\documentclass[12pt]{article}
\usepackage{amssymb,amsmath,epsfig}

\begin{document}

\title{\bf Expansionfree Fluid Evolution and Skripkin Model in $f(R)$ Theory}

\author{M. Sharif \thanks{msharif.math@pu.edu.pk} and H. Rizwana
Kausar
\thanks{rizwa\_math@yahoo.com}\\
Department of Mathematics, University of the Punjab,\\
Quaid-e-Azam Campus, Lahore-54590, Pakistan.}

\date{}
\maketitle

\begin{abstract}
We consider the modified $f(R)$ theory of gravity whose higher
order curvature terms are interpreted as a gravitational fluid or
dark source. The gravitational collapse of a spherically symmetric
star, made up of locally anisotropic viscous fluid, is studied
under the general influence of the curvature fluid. Dynamical
equations and junction conditions are modified in the context of
$f(R)$ dark energy and by taking into account the expansionfree
evolution of the self-gravitating fluid. As a particular example,
the Skripkin model is investigated which corresponds to isotropic
pressure with constant energy density. The results are compared
with corresponding results in General Relativity.
\end{abstract}

\section{Introduction}

Modified $f(R)$ theory of gravity constitutes an important
development in modern cosmology and theoretical physics. One of the
main motivations of this theory is that it may lead to some
interesting results about dark energy (DE). The other motivation
comes from the fact that every unification of fundamental
interaction exhibits effective actions containing higher order terms
in the curvature invariants. This strategy was adopted in the study
of quantum field theory in curved spacetimes \cite{1} as well as in
the Lagrangian of string and Kaluza-Klein theories \cite{2}.

Higher order terms always give an even number as an order of the
field equations. For example, $R^2$ term produces fourth order field
equations \cite{3}, term $R\Box R$ (where $\Box \equiv
\nabla^{\mu}\nabla_{\mu}$) gives sixth order field equations
\cite{4,5}, similarly, $R\Box^2 R$ yields eights order field
equations \cite{6} and so on. Using conformal transformation, the
term with second derivative corresponds to a scalar field. For
instance, fourth order gravitational theory corresponds to Einstein
theory with one scalar field, sixth order gravity corresponds to
Einstein gravity with two scalar fields, \textit{etc.} \cite{4,7}.
In this context, it is easy to show that $f(R)$ gravity is
equivalent to scalar tensor theory as well as to General Relativity
(GR) with an ideal fluid \cite{8}.

Let us now show that how $f(R)$ gravity can be related to the
problem of DE by a straightforward argument. When the
Einstein-Hilbert (EH) gravitational action in GR,
\begin{equation}\label{1}
S_{EH}=\frac{1}{2 \kappa}\int d^{4}x\sqrt{-g}R,
\end{equation}
is written in the modified form as follows
\begin{equation}\label{2}
S_{f(R)}=\frac{1}{2\kappa}\int d^{4}x\sqrt{-g}f(R),
\end{equation}
the addition of a non-linear function of the Ricci scalar
demonstrates to cause acceleration for a wide variety of $f(R)$
function, e.g., \cite{fr1}-\cite{fr10}. Variation of $f(R)$ action
with respect to the metric tensor leads to the following fourth
order partial differential equations
\begin{equation}\label{3}
F(R)R_{\alpha\beta}-\frac{1}{2}f(R)g_{\alpha\beta}-\nabla_{\alpha}
\nabla_{\beta}F(R)+ g_{\alpha\beta} \Box F(R)=\kappa
T_{\alpha\beta},\quad(\alpha,\beta=0,1,2,3),
\end{equation}
where $F(R)\equiv df(R)/dR$. Writing this equation in the form of
Einstein tensor, it follows that \cite{fr11}
\begin{equation}\label{4}
G_{\alpha\beta}=\frac{\kappa}{F}(T_{\alpha\beta}^{m}+T_{\alpha\beta}^{(D)}),
\end{equation}
where
\begin{equation}\label{5}
T_{\alpha\beta}^{(D)}=\frac{1}{\kappa}\left[\frac{f(R)+R
F(R)}{2}g_{\alpha\beta}+\nabla_{\alpha} \nabla_{\beta}F(R)
-g_{\alpha\beta} \Box F(R)\right].
\end{equation}
It is clear from Eq.(\ref{4}) that curvature stress-energy tensor
$T_{\alpha\beta}^{(D)}$ formally plays the role of a source in the
field equations and its effect is the same as that of an effective
fluid of purely geometrical origin. In fact, this scheme provides
all the ingredients needed to tackle the dark side of the universe.
Thus curvature fluid can play the role of both dark matter and DE.
The "Dark source" term is not restricted to hold the usual energy
conditions. Therefore, $f(R)$ theory may be used to explain the
effects of DE on cosmological and gravitational phenomena of the
universe.

The importance of gravitational collapse lies at the center of
structure formation in the universe. A starting smooth arrangement
of matter will eventually collapse and make the powerful
structures such as stellar groups, planets and stars. One can
study the gravitational collapse by taking the interior and
exterior regions of spacetime. The proper junction conditions help
to study the smooth matching of these regions. In Einstein
gravity, the pioneer work \cite{OS} was carried out on dust
collapse by taking static Schwarzschild in the exterior and
Friedmann like solution in the interior spacetime.

During fluid evolution, self-gravitating objects may pass through
phases of intense dynamical activities. The dynamical equations
are used to observe the effects of dissipation over collapsing
process. Skripkin \cite{skrip} studied the central explosion of
the spherically symmetric fluid distribution under condition of
constant energy density. The evolution of such a fluid yields the
formation of a Minkowskian cavity within the fluid distribution
centered at the origin. This problem is studied in detail by
Herrera \emph{et al.} \cite{H1,H2}. It was shown that under
Skripkin conditions, the expansion scalar vanishes which requires
the existence of a cavity within the fluid. Sharif \emph{et al.}
investigated some aspects of gravitational collapse regarding
singularity and event horizons in $f(R)$ theory \cite{we1} and
also, in GR dissipative fluid collapse with and without adding
charge \cite{szak}-\cite{sa}. In a recent paper \cite{we2}, we
have studied the effects of $f(R)$ DE on a dissipative collapse.
Some special solutions are also discussed for nondissipative case.

In this work, we investigate that how DE generated by curvature
fluid generally affects the dynamics of viscous self-gravitating
fluid. In fact, this provides the generalization and extension of
our previous work \cite{we2}. We explore the contribution of
$f(R)$ DE in developing Skripkin model. The plan of the paper is
as follows. In next section \textbf{2}, we formulate the field
equations in metric $f(R)$ gravity for spherically symmetric
distribution of anisotropic fluid. Section \textbf{3} is devoted
to derive junction conditions and dynamical equations with the
inclusion of curvature fluid. In section \textbf{4}, fluid
evolution is investigated under the assumption of vanishing scalar
expansion. Section \textbf{5} is devoted to study the Skripkin
model in $f(R)$ gravity. The last section \textbf{6} concludes the
main results of the paper.

\section{Fluid Distribution and the Field Equations}

We consider a $3D$ hypersurface $\Sigma^{(e)}$, an external
boundary of the star, which divides a $4D$ spherically symmetric
spacetime into two regions named as interior and exterior
spacetimes. For the interior region to $\Sigma^{(e)}$, we take the
line element in the form as
\begin{equation}\setcounter{equation}{1}\label{6}
ds^2_-=A^2(t,r)dt^{2}-B^2(t,r)dr^{2}-C^2(t,r)(d\theta^{2}+\sin^2\theta
d\phi^{2}).
\end{equation}
For the exterior spacetime to $\Sigma^{(e)}$, we take the
Schwarzschild spacetime given by the line element
\begin{equation}\label{7}
ds^2_+=(1-\frac{2m}{r})d\nu^2+2drd\nu-r^2(d\theta^2+\sin^2\theta
d\phi^2),
\end{equation}
where $m$ represents the total mass and $\nu$ is the retarded
time.

In the interior region, we assume a distribution of anisotropic
self-gravitating fluid which undergoes dissipation in the form of
shear viscosity. The energy-momentum tensor for such a fluid is
defined as
\begin{equation}\label{8}
T_{\alpha\beta}=(\rho+p_{\perp})u_{\alpha}u_{\beta}-p_{\perp}g_{\alpha\beta}+
(p_r-p_{\perp})\chi_{\alpha}\chi_{\beta}-2\eta\sigma_{\alpha\beta}.
\end{equation}
Here, we have $\rho$ as the energy density, $p_{\perp}$ the
tangential pressure, $p_r$ the radial pressure,  $\eta$ the
coefficient of shear viscosity, $u_{\alpha}$ the four-velocity of
the fluid and $\chi_{\alpha}$ the unit four-vector along the
radial direction. These quantities satisfy the relations
\begin{equation}\label{9}
u^{\alpha}u_{\alpha}=1,\quad\chi^{\alpha}\chi_{\alpha}=-1,\quad
\chi^{\alpha}u_{\alpha}=0
\end{equation}
which are obtained from the following definitions in co-moving
coordinates
\begin{equation}\label{10}
u^{\alpha}=A^{-1}\delta^{\alpha}_{0},\quad
\chi^{\alpha}=B^{-1}\delta^{\alpha}_{1}.
\end{equation}
The shear tensor $\sigma_{ab}$ is defined by
\begin{equation}\label{11}
\sigma_{\alpha\beta}=u_{(\alpha;\beta)}-a_{(\alpha}u_{\beta)}-\frac{1}{3}\Theta
(g_{\alpha\beta}-u_{\alpha}u_{\beta}),
\end{equation}
where the acceleration $a_{a}$ and the expansion scalar $\Theta$ are
given by
\begin{equation}\label{12}
a_{\alpha}=u_{\alpha;\beta}u^{\beta},\quad
\Theta=u^{\alpha}_{;\alpha}.
\end{equation}
The bulk viscosity does not appear as it can be absorbed in the
form of radial and tangential pressures of the self-gravitating
fluid. From Eqs.(\ref{10}) and (\ref{11}), we have the following
non-vanishing components of the shear tensor
\begin{equation}\label{13}
\sigma_{11}=-\frac{2}{3}{B^{2}}\sigma,\quad
\sigma_{22}=\frac{1}{3}{C^{2}}\sigma,\quad
\sigma_{33}=\sigma_{22}\sin^2\theta,
\end{equation}
where $\sigma$ is the shear scalar and is given by
\begin{equation}\label{14}
\sigma=\frac{1}{A}\left(\frac{\dot{B}}{B}-\frac{\dot{C}}{C}\right).
\end{equation}
From Eqs.(\ref{10}) and (\ref{12}), we have
\begin{equation}\label{15}
a_{1}=-\frac{A'}{A},\quad a^2=a^\alpha
a_\alpha=\left(\frac{A'}{AB}\right)^2,\quad
\Theta=\frac{1}{A}\left(\frac{\dot{B}}{B}+2\frac{\dot{C}}{C
}\right),
\end{equation}
where dot and prime represent derivative with respect to $t$ and $r$
respectively.

In Einstein frame, the field equations (\ref{4}) for the interior
metric become
\begin{eqnarray}\nonumber
&&\left(\frac{2\dot{B}}{B}+\frac{\dot{C}}{C}\right)
\frac{\dot{C}}{C}-\left(\frac{A}{B}\right)^2
\left[\frac{2C''}{C}+\left(\frac{C'}{C}\right)^2-\frac{2B'C'}{BC}
-\left(\frac{B}{C}\right)^2\right]\\\label{16}
&&=\frac{8\pi}{F}[{\rho}A^{2}+T^{(D)}_{00}], \\\label{17}
&&-2\left(\frac{\dot{C'}}{C}-\frac{\dot{C}A'}{CA}-\frac{\dot{B}C'}{BC}\right)
=\frac{8\pi}{F}T^{(D)}_{01},\\\nonumber
&&-\left(\frac{B}{A}\right)^2\left[\frac{2\ddot{C}}{C}-\left(\frac{2\dot{A}}{A}
-\frac{\dot{C}}{C}\right) \frac{\dot{C}}{C}\right]
+\left(\frac{2A'}{A}+\frac{C'}{C}\right)\frac{C'}{C}-\left(\frac{B}{C}\right)^2
\end{eqnarray}
\begin{eqnarray}
\label{18}
&&=\frac{8\pi}{F}[(p_r+\frac{4}{3}{\eta}{\sigma})B^{2}+T^{(D)}_{11}],\\\nonumber
&&-\left(\frac{C}{A}\right)^2\left[\frac{\ddot{B}}{B}-\frac{\ddot{C}}{C}-\frac{\dot{A}}{A}
\left(\frac{\dot{B}}{B}+\frac{\dot{C}}{C}\right)+\frac{\dot{B}\dot{C}}{BC}\right]
+\left(\frac{C}{B}\right)^2\left[\frac{A''}{A}+\frac{C''}{C}\right.\\\label{19}
&&\left.-\frac{A'B'}{AB}+\left(\frac{A'}{A}-\frac{B'}{B}\right)\frac{C'}{C}\right]
=\frac{8\pi}{F}[(p_{\perp}-\frac{2}{3}{\eta}{\sigma})C^2+T^{(D)}_{22}].
\end{eqnarray}
Here the components of dark fluid are obtained from Eq.(\ref{5})
as follows
\begin{eqnarray}\nonumber
T_{00}^{(D)}&=&\frac{A^2}{8\pi}\left[\frac{f+R
F}{2}+\frac{F''}{B^2}+\left(\frac{2\dot{C}}{C}-\frac{\dot{B}}{B}\right)
\frac{\dot{F}}{A^2}+\left(\frac{2C'}{C}-\frac{B'}{B}\right)\frac{F'}{B^2}\right],\\\nonumber
T_{01}^{(D)}&=&\frac{1}{8\pi}\left(\dot{F}'
-\frac{A'}{A}\dot{F}-\frac{\dot{B}}{B}F'\right),\\\nonumber
T_{11}^{(D)}&=&\frac{-B^2}{8\pi}\left[\frac{f+R
F}{2}-\frac{\ddot{F}}{A^2}+\left(\frac{\dot{A}}{A}+\frac{2\dot{C}}{C}\right)
\frac{\dot{F}}{A^2}+\left(\frac{A'}{A}+\frac{2C'}{C}\right)\frac{F'}{B^2}\right],\\\nonumber
T_{22}^{(D)}&=&\frac{-C^2}{8\pi}\left[\frac{f+R
F}{2}-\frac{\ddot{F}}{A^2}+\frac{F''}{B^2}
+\left(\frac{\dot{A}}{A}-\frac{\dot{B}}{B}+\frac{\dot{C}}{C}\right)
\frac{\dot{F}}{A^2}\right.\\\label{t22}
&+&\left.\left(\frac{A'}{A}-\frac{B'}{B}+\frac{C'}{C}\right)\frac{F'}{B^2}\right].
\end{eqnarray}
The Ricci scalar curvature is given by
\begin{eqnarray}\nonumber
R&=&2\left[\frac{A''}{AB^2}-\frac{\ddot{B}}{A^2B}+\frac{\dot{A}\dot{B}}{A^3B}
-\frac{A'B'}{AB^3}+\frac{2\ddot{C}}{CA^2}+\frac{2\dot{A}\dot{C}}{ACB^2}\right.\\\label{R}
&+&\left.\frac{2C''}{CB^2}-\frac{2\dot{C}\dot{B}}{CA^2B}
-\frac{2C'B'}{CB^3}-\frac{1}{C^2}-\frac{\dot{C}^2}{A^2C^2}+\frac{C'^2}{B^2C^2}\right].
\end{eqnarray}

\section{Junction Conditions and the Dynamical Equations}

Here, we develop equations that govern the dynamics of dissipative
spherically symmetric collapsing process. For this purpose, we use
Misner and Sharp formalism \cite{MS}. The mass function is defined
by
\begin{equation}\label{20}
M(t,r)=\frac{C}{2}(1+g^{\mu\nu}C_{,\mu}C_{,\nu})=\frac{C}{2}\left(1+\frac{\dot{C}^2}{A^2}
-\frac{C'^2}{B^2}\right).
\end{equation}
From the continuity of the first and second differential forms,
the matching of the nonadiabatic sphere to the Schwarzschild
spacetime on the boundary surface, ${\Sigma^{(e)}}$, yields the
following results
\begin{equation}\label{j1}
M(t,r)\overset{\Sigma^{(e)}}{=}m
\end{equation}
and
\begin{eqnarray}\nonumber
&&2\left(\frac{\dot{C'}}{C}-\frac{\dot{C}A'}{CA}-\frac{\dot{B}C'}{BC}\right)
\overset{\Sigma^{(e)}}{=}
-\frac{B}{A}\left[\frac{2\ddot{C}}{C}-\left(\frac{2\dot{A}}{A}-
\frac{\dot{C}}{C}\right) \frac{\dot{C}}{C}\right]\\\label{j2} &&
+\frac{A}{B}\left[\left(\frac{2A'}{A}+\frac{C'}{C}\right)\frac{C'}{C}
-\left(\frac{B}{C}\right)^2\right].
\end{eqnarray}
The important reason for generating the junction conditions at the
stellar surface is to study the dissipative evolution of the star.
Using the field equations (\ref{17}) and (\ref{18}) in
Eq.(\ref{j2}), we obtain
\begin{equation}\label{j3}
-p_r-\frac{4}{3}\eta\sigma\overset{\Sigma^{(e)}}{=}\frac{T^{(D)}_{11}}{B^2}
+\frac{T^{(D)}_{01}}{AB}.
\end{equation}
In the next section, we would discuss dynamics with physically
meaningful assumption of vanishing expansion scalar which
describes the rate of change of small volumes of the fluid. The
expansionfree fluid evolution should imply the formation of a
vacuum cavity within spherically symmetric fluid distribution
\cite{H1}. Taking ${\Sigma^{(i)}}$ ($i$ stands for internal) to be
the boundary surface of that vacuum cavity and matching this
surface with Minkowski spacetime, we have
\begin{equation}\label{j4}
M(t,r)\overset{\Sigma^{(i)}}{=}0, \quad
-p_r-\frac{4}{3}\eta\sigma\overset{\Sigma^{(i)}}{=}\frac{T^{(D)}_{11}}{B^2}
+\frac{T^{(D)}_{01}}{AB}.
\end{equation}

The proper time and radial derivatives are given by
\begin{equation}\label{45}
D_{T}=\frac{1}{A}\frac{\partial}{\partial t},\quad
D_{C}=\frac{1}{C'}\frac{\partial}{\partial r},
\end{equation}
where $C$ is the areal radius of a spherical surface inside the
boundary $\Sigma^{(e)}$, as measured from its area. The velocity of
the collapsing fluid is defined by the proper time derivative of
$C$, i.e.,
\begin{equation}\label{22}
U=D_{T}C=\frac{\dot{C}}{A}
\end{equation}
which is always negative. Using this expression, Eq.(\ref{20})
implies that
\begin{equation}\label{23}
E\equiv\frac{C'}{B}=[1+U^{2}+\frac{2M}{C}]^{1/2}.
\end{equation}
When we make use of Eqs.(\ref{14}), (\ref{15}) and (\ref{45}) in
Eq.(\ref{17}), we obtain
\begin{equation}\label{24}
E\left[\frac{1}{3}D_{C}(\Theta-\sigma)-\frac{\sigma}{C}\right]
=-\frac{4\pi}{F}\frac{T^{(D)}_{01}}{AB}.
\end{equation}

The rate of change of mass in Eq.(\ref{20}) with respect to proper
time, with the use of Eqs.(\ref{16})-(\ref{19}), is given by
\begin{equation}\label{25}
D_{T}M=\frac{-4\pi}{F}\left[\left(p_r+\frac{4}{3}{\eta}{\sigma}
+\frac{T^{(D)}_{11}}{B^2}\right)U
-E\frac{T^{(D)}_{01}}{AB}\right]C^2.
\end{equation}
This represents variation of total energy inside a collapsing
surface of radius $C$. In the case of collapse $U<0$, the terms
inside the first round brackets increases the energy density
through the rate of work being done by the effective radial
pressure $p_r+\frac{4}{3}{\eta}{\sigma}$. The presence of dark
fluid component shows the contribution of DE having large negative
pressure. These terms appear with positive sign representing
negative effect, hence decrease the rate of change of mass with
respect to time. Now, it depends upon the strength of DE terms
that they may balance the positive effect of effective radial
pressure or overcome on them. Likewise, we have
\begin{equation}\label{26}
D_{C}M=\frac{4\pi}{F}\left[\rho+\frac{T^{(D)}_{00}}{A^2}
-\frac{U}{E}\frac{T^{(D)}_{01}}{AB}\right]C^2.
\end{equation}
This equation describes how energy density and curvature terms
influence the mass between neighboring surfaces of radius $C$ in the
fluid distribution. Here the rate would decrease in the consecutive
surfaces by the repulsive effect of DE. Integration of Eq.(\ref{26})
with respect to "$C$" leads to
\begin{equation}\label{27}
M=4\pi\int^{C}_{0}\frac{C^2}{2F}\left[\rho+\frac{T^{(D)}_{00}}{A^2}
-\frac{U}{E}\frac{T^{(D)}_{01}}{AB}\right]dC.
\end{equation}

The dynamical equations can be obtained from the contracted Bianchi
identities. Consider the following two equations
\begin{eqnarray}\label{52}
\left(T^{\alpha\beta}+\overset{(D)}{T^{\alpha\beta}}\right)_{;\beta}u_{\alpha}=0,\quad
\left(T^{\alpha\beta}+\overset{(D)}{T^{\alpha\beta}}\right)_{;\beta}
\chi_{\alpha}=0
\end{eqnarray}
which yield
\begin{eqnarray}\label{28}
\overset{(D)}{T^{\alpha\beta}}_{;\beta}u_{\alpha}&=&
-\frac{1}{A}\left[\dot{\rho}+(\rho+p_r+\frac{4}{3}\eta
\sigma)\frac{\dot{B}}{B}+2(\rho+p_{\perp}-\frac{2}{3}\eta
\sigma)\frac{\dot{C}}{C}\right],\\\nonumber
\overset{(D)}{T^{\alpha\beta}}_{;\beta}
\chi_{\alpha}&=&\frac{1}{B}\left[(p_r+\frac{4}{3}\eta\sigma)'+(\rho+p_r+\frac{4}{3}\eta
\sigma)\frac{A'}{A}\right.\\\label{29}&+&\left.2(p_r-p_{\perp}+2\eta
\sigma)\frac{C'}{C}\right].
\end{eqnarray}
Using Eqs.(\ref{14}), (\ref{15}), (\ref{45}) and (\ref{23}), it
follows that
\begin{eqnarray}\label{30}
\overset{(D)}{T^{\alpha\beta}}_{;\beta}u_{\alpha}&=&
-[D_T\rho+\frac{1}{3}(3\rho+p_r+2p_{\perp})\Theta+\frac{2}{3}
(p_r-p_{\perp}-2\eta \sigma)\sigma],\\\nonumber
\overset{(D)}{T^{\alpha\beta}}_{;\beta}
\chi_{\alpha}&=&ED_C(p_r+\frac{4}{3}\eta\sigma)+(\rho+p_r+\frac{4}{3}\eta
\sigma)a+2(p_r-p_{\perp}+2\eta \sigma)\frac{E}{C}.\\\label{31}
\end{eqnarray}
The acceleration $D_{T}U$ of the collapsing matter inside the
hypersurface is obtained by using Eqs.(\ref{18}), (\ref{45}) and
(\ref{23}) as follows
\begin{eqnarray}\label{32}
D_{T}U&=&-\frac{M}{C^2}-\frac{4\pi}{F}\left(p_r+\frac{4}{3}\eta\sigma
+\frac{T^{(D)}_{11}}{B^2}\right)C+Ea.
\end{eqnarray}
Substituting $a$ from Eq.(\ref{32}) into (\ref{31}), it follows
that
\begin{eqnarray}\nonumber
&&(\rho+p_r+\frac{4}{3}\eta\sigma)D_TU\\\nonumber
&=&-(\rho+p_r+\frac{4}{3}\eta
\sigma)\left[\frac{M}{C^2}+\frac{4\pi}{F}\left(p_r+\frac{4}{3}\eta\sigma
+\frac{T^{(D)}_{11}}{B^2}\right)C\right]\\\label{38'}
&-&E^{2}\left[D_C(p_r+\frac{4}{3}\eta
\sigma)+\frac{2}{C}(p_r-p_{\perp}+2\eta \sigma)\right]
\end{eqnarray}
This equation shows the role of different forces on the collapsing
process. The factor within the brackets on the left hand side
stands for "effective" inertial mass and the remaining term is
acceleration. The first term on the right hand side represents
gravitational force. The term within the first square brackets
shows how shear viscosity and DE affect the passive gravitational
mass. The first two terms in the second square brackets are
gradient of the effective pressure and effect of local anisotropy
of pressure with negative sign which increases the rate of
collapse.

\section{Expansionfree Evolution of Self-gravitating Fluid}

In this section, we take $\Theta=0$ and discuss expansionfree
evolution of the self-gravitating fluid. In such a case,
Eq.(\ref{15}) yields
\begin{equation}\setcounter{equation}{1}\label{35}
\frac{\dot{B}}{B}=-2\frac{\dot{C}}{C}\quad \Rightarrow\quad
B=\frac{g_1(r)}{C^2},
\end{equation}
where $g_1(r)$ is an arbitrary function. Implication of vanishing
$\Theta$ requires that innermost shell of the fluid should be away
from the center of the collapsing sphere. This situation initiates
the formation of a vacuum cavity at the center \cite{H1}. The
physical meaning of this condition is explained by taking two
different definitions of radial velocity of the fluid.
Substituting Eq.(4.1) and value of $T^{(D)}_{01}$ from
Eq.(\ref{t22}) in (\ref{17}), we get
\begin{equation}\label{36}
2\left(\frac{\dot{C'}}{\dot{C}}-\frac{A'}{A}-2\frac{C'}{C}\right)+\left(\frac{\dot{F'}}{
F}-\frac{A'\dot{F}}{AF}\right)\frac{C}{\dot{C}}+2\frac{F'}{F}=0.
\end{equation}
Integration yields
\begin{equation}\label{37}
A=\frac{F^2\dot{C}C^2}{\tau(t)}e^{-\int{(\frac{\dot{F'}}{
F}-\frac{A'\dot{F}}{AF})\frac{C}{\dot{C}}dr}}.
\end{equation}
Thus the interior metric becomes
\begin{eqnarray}\nonumber
ds^2&=&\left(\frac{F^2\dot{C}C^2}{\tau_1(t)}\exp\left[{-\int{(\frac{\dot{F'}}{
F}-\frac{A'\dot{F}}{AF})\frac{C}{\dot{C}}dr}}\right]\right)^2dt^2
\\\label{38}&&-\left(\frac{g_1}{C^2}\right)^2dr^2-C^2(t,r)(d\theta^{2}+\sin^2\theta
d\phi^{2}).
\end{eqnarray}
The above metric represents spherically symmetric anisotropic
fluid which is going under shearing expansionfree evolution.

To proceed further, it is necessary to assume the condition of
constant scalar curvature ($R=R_{c}$), according to which
$F(R_c)=constant$. In $f(R)$ theory, the case of constant scalar
curvature exhibits behavior just like solutions with cosmological
constant in GR. This is one of the reason why the DE issue can be
addressed by using this theory. In this case metric reduce to the
following
\begin{equation}\label{44}
ds^2=\left(\frac{F_c^2\dot{C}C^2}{\tau_1(t)}\right)^2dt^2-\frac{1}{C^4}dr^2
-C^2(t,r)(d\theta^{2}+\sin^2\theta d\phi^{2}).
\end{equation}
Here we have chosen $g_1(r)=1$ without loss of generality. For this
metric, the scalar curvature (\ref{R}) becomes
\begin{eqnarray}\nonumber
R&=&2\left[\left(\frac{\dot{C}''C^4}{\dot{C}}+2C''C^3+\frac{4\dot{C}'C'C^3}{\dot{C}}
+2C'^2C^2+\frac{2\dot{C}'C'C^3}{\dot{C}}
+9C'^2C^2\right.\right.\\\nonumber
&+&\left.\left.4\dot{C}^2C^2+2\ddot{C}C^3-\frac{2\dot{\tau}\dot{C}C^3}{\tau}
+2C''C^3+\frac{1}{C^2}\right)+\frac{\tau_1^2}{F_c^4}\left(\frac{\ddot{C}}{C}
-\frac{6\dot{C}^2}{C^2}\right.\right.\\\label{Rc}
&-&\left.\left.\frac{2\ddot{C}}{\dot{C}^2C^5}+\frac{2\dot{\tau}}{\tau
C^5\dot{C}}-\frac{2\ddot{C}}{\dot{C}C^3}+\frac{7}{C^6}\right)\right].
\end{eqnarray}
One can find such values of $C$ and $\tau_1$ for which the condition
of constant scalar curvature is satisfied. For constant scalar
curvature, the components of dark fluid (\ref{t22}) reduce to
\begin{equation}\label{44'}
T_{00}^{(D_c)}=\frac{A^2\Omega}{8\pi},\quad T_{01}^{(D_c)}=0,\quad
T_{11}^{(D_c)}=T_{22}^{(D_c)}=\frac{-C^2\Omega}{8\pi},
\end{equation}
where
\begin{equation}
\Omega=\frac{f(R_c)+R_c F(R_c)}{2}.
\end{equation}
The corresponding junction conditions (\ref{j1}), (\ref{j3}) and
(\ref{j4}) will take the form
\begin{eqnarray}\nonumber
&&M(t,r)\overset{\Sigma^{(e)}}{=}m\quad\quad
p_r\overset{\Sigma^{(e)}}{=}\frac{\Omega}{8\pi},\\\label{j1'}
&&M(t,r)\overset{\Sigma^{(i)}}{=}0, \quad\quad
p_r\overset{\Sigma^{(i)}}{=}\frac{\Omega}{8\pi}.
\end{eqnarray}
Using Eq.(\ref{44}) in the field equations (\ref{16})-(\ref{19}),
it follows that
\begin{eqnarray}\label{76}
&&-2C^3C''-5C^2C'^2+\frac{1}{C^2}-3\frac{\tau_1^2}{F_c^4C^6}
=\frac{1}{F_c}(8\pi\rho+\Omega),\\\label{77}
&&\frac{1}{3}D_C\sigma+\frac{\sigma}{C}=0,\\\nonumber
&&\frac{\tau_1^2}{F_c^4\dot{C}C^5}\left(3\frac{\dot{C}}{C}-2
\frac{\dot{\tau_1}}{\tau_1}\right)+
C^3C'\left(2\frac{\dot{C}'}{\dot{C}}+5\frac{C'}{C}\right)-\frac{1}{C^2}\\\label{78}
&&=\frac{1}{F_c}(8\pi p_r-\Omega),\\\nonumber
&&\frac{-\tau_1^2}{F_c^4\dot{C}C^5}\left(6\frac{\dot{C}}{C}-
\frac{\dot{\tau_1}}{\tau_1}\right)-
C^4\left(\frac{\dot{C}''}{\dot{C}}+7\frac{\dot{C}'C'}{\dot{C}C}+3\frac{C''}{C}
+10\frac{C'^2}{C^2}\right)\\\label{79} &&=\frac{1}{F_c}(8\pi
p_{\perp}-\Omega).
\end{eqnarray}
Similarly, Eqs.(\ref{25})-(\ref{26}) become
\begin{eqnarray}\label{25'}
D_{T}M&=&\frac{-4\pi
U}{F_c}\left(p_r-\frac{\Omega}{8\pi}\right)C^2,\\\label{26'}
D_{C}M&=&\frac{4\pi}{F_c}\left(\rho+\frac{\Omega}{8\pi}\right)C^2
\end{eqnarray}
implying that
\begin{equation}\label{27'}
M=\frac{4\pi}{F_c}\int^{C}_{0}\left(\rho+\frac{\Omega}{8\pi}\right)C^2dC.
\end{equation}
Further, Eq.(\ref{25'}) yields
\begin{equation}\label{251}
8\pi p_r=\frac{-2\dot{M}F_c}{C^2\dot{C}}+\Omega.
\end{equation}
For Schwarzschild mass $m$, Eq.(\ref{251}) is fully consistent with
the junction conditions Eq.(\ref{j1'}).

On the similar conditions, Bianchi identities are reduced to the
following
\begin{eqnarray}\label{53}
&&\dot{\rho}+2(p_{\perp}-p_r)\frac{\dot{C}}{C}+\frac{\Omega}{16\pi}
\frac{\tau_1(2C\dot{C}^2+C^2\ddot{C})-\dot{\tau_1}\dot{C}C^2}{\tau_1C^2\dot{C}}=0,
\\\label{56}
&&p_r'+(\rho+p_r)\frac{\dot{C}'}{\dot{C}}+2(\rho+2p_r-p_{\perp})\frac{C'}{C}
-\frac{\Omega}{\pi}\frac{C'}{C}=0.
\end{eqnarray}
We would like to mention here that in GR for isotropic fluid,
i.e., $p_r=p_{\perp}$, Eq.(\ref{53}) implies that energy density
$\rho$ depends only on $r$. However, in $f(R)$ theory it remains
the function of both time and radial coordinate. Integration of
Eq.(\ref{76}) after some manipulation yields
\begin{equation}\label{82}
C'^2=\frac{1}{C^4}+\frac{\tau_2-2m}{C^5}+\frac{\tau_1^2}{F_c^4}\frac{1}{C^8}
-\frac{\Omega}{C^2} \left(\frac{1}{6\pi}-\frac{1}{F_c}\right).
\end{equation}
Here $\tau_2(t)$ is an arbitrary integration function. Using
Eqs.(\ref{44}) and (\ref{82}) into Eq.(\ref{20}), we obtain
\begin{equation}\label{82'}
\tau_2=-\frac{\Omega}{k}\left(\frac{1}{6\pi}-\frac{1}{F_c}\right).
\end{equation}

\section{Skripkin Model}

In the Skripkin model, it is assumed that fluid has isotropic
pressure ($p_r=p_\perp=p$) and constant energy density, i.e.,
$\rho=\rho_0$. Here the fluid is assumed to be at rest initially
and after that there is a sudden explosion at the center keeping
the Skripkin condition. It is noted here that in GR, the expansion
scalar automatically vanishes for the Skripkin conditions by
virtue of Eq.(\ref{30}). However, in $f(R)$ gravity, it will not
vanish and we take expansionfree evolution in order to study
Skripkin model. Consequently, Eq.(\ref{82}) becomes
\begin{equation}\label{83}
C'^2=\frac{1}{C^4}+\frac{\tau_2}{C^5}+\frac{\tau_1}{F_c^4C^8}-\frac{k}{C^2}
-\frac{\Omega}{F_c}\frac{1}{C^2},
\end{equation}
where
\begin{equation}\label{84}
k=\frac{8\pi\rho_0F_c}{3}.
\end{equation}
Under Skripkin conditions, Eq.(\ref{38'}) takes the form
\begin{equation}\label{85}
(\rho_0+p)D_TU=-(\rho_0+p)\left[\frac{m}{C^2}+4\pi pC-\frac{\Omega
C}{2}\right]-E^2D_Cp,
\end{equation}
Using Eq.(\ref{83}) in Eq.(\ref{78}), we obtain
\begin{equation}\label{86}
\frac{8\pi
p}{F_c}=-\left(3k+2\frac{\Omega}{F_c}\right)+\frac{\dot{\tau_2}}{C^2\dot{C}}.
\end{equation}

For isotropic pressure, Eq.(\ref{j1'}) becomes
\begin{equation}\label{j1''}
p\overset{\Sigma^{(e)}}{=}\frac{\Omega}{8\pi}.
\end{equation}
Substituting Eq.(\ref{j1''}) in Eq.(\ref{86}) and integrating, we
get
\begin{equation}\label{88}
\tau_2=\left(k+\frac{\Omega}{3F_c}\right)C^3_{\Sigma^{(e)}}+c_1,
\end{equation}
where $c_1$ is an integration constant. Using Eqs.(\ref{44}) and
(\ref{83}) into Eq.(\ref{20}), the mass function becomes
\begin{equation}\label{89}
M=\frac{k}{2}(C^3-C^3_{\Sigma^{(e)}})+\frac{\Omega}{2F_c}(C^3-
\frac{1}{3}C^3_{\Sigma^{(e)}}) -\frac{c_1}{2}.
\end{equation}
The total mass of the configuration $m$ is obtained by measuring
$M$ on the boundary ${\Sigma^{(e)}}$ as follows
\begin{equation}\label{90}
M=\frac{\Omega}{3F_c}C^3_{\Sigma^{(e)}}-\frac{c_1}{2}=m,
\end{equation}
Substituting value of $c_1$ from Eq.(\ref{90}) into Eqs.(\ref{88})
and (\ref{89}), we have
\begin{equation}\label{91}
\tau_2=\left(k+\frac{\Omega}{F_c}\right)C^3_{\Sigma^{(e)}}-2m,
\end{equation}
\begin{equation}\label{92}
M=\frac{1}{2}\left(k+\frac{\Omega}{F_c}\right)(C^3-C^3_{\Sigma^{(e)}})+m,
\end{equation}
Applying the matching conditions (\ref{j4}) on the boundary
surface of the cavity inside the fluid, Eq.(\ref{92}) yields
\begin{equation}\label{93}
m=\frac{1}{2}(k+\frac{\Omega}{F_c})(C^3_{\Sigma^{(e)}}-C^3_{\Sigma^{(i)}}).
\end{equation}
Using this equation in Eq.(\ref{91}), we obtain
\begin{equation}\label{94}
\tau_2=\left(k+\frac{\Omega}{F_c}\right)C^3_{\Sigma^{(i)}}.
\end{equation}
Differentiating Eq.(\ref{93}) with respect to "t", we have
\begin{equation}\label{95}
C^2_{\Sigma^{(e)}}\dot{C}_{\Sigma^{(e)}}=C^2_{\Sigma^{(i)}}\dot{C}_{\Sigma^{(i)}}
\end{equation}
implying that
\begin{equation}\label{96,97}
A_{\Sigma^{(e)}}=A_{\Sigma^{(i)}}\quad \textmd{and}\quad
U_{\Sigma^{(e)}}=\frac{C^2_{\Sigma^{(i)}}}{C^2_{\Sigma^{(e)}}}U_{\Sigma^{(i)}}.
\end{equation}
This result can also be obtained by definition of $U$ on the
boundary surfaces.

Applying Skripkin condition of constant energy density,
Eq.(\ref{27'}) yields
\begin{equation}\label{27''}
M=\frac{4\pi}{F_c}\left(\rho+\frac{\Omega}{8\pi}\right)\frac{C^3}{3}.
\end{equation}
Differentiating with respect to "t", we have
\begin{equation}\label{27'''}
\dot{M}=\frac{4\pi}{F_c}\left(\rho+\frac{\Omega}{8\pi}\right)C^2\dot{C}.
\end{equation}
Substitution of Eq.(\ref{27'''}) into Eq.(\ref{25'}) with
isotropic pressure yields
\begin{equation}\label{250}
p=-\frac{\rho_0}{F_c}+\frac{\Omega}{8\pi}\left(1-\frac{1}{F_c}\right)=\textmd{constant}.
\end{equation}
Comparing Eq.(\ref{250}) with Eq.(\ref{j1''}), we obtain
\begin{equation}\label{252}
\rho_0\overset{\Sigma^{(e)}}{=}\frac{\Omega}{8\pi}.
\end{equation}
This shows that Skripkin model is eliminated by the junction
conditions. It is mentioned here that in GR, $\rho_0=0$ while in
our case, it turns out to be a non-zero constant.

Since pressure will be maximum on some spherical surface $C=C_s$
(say), hence pressure gradient must vanish on this surface. Thus it
follows from Eq.(\ref{56}) that
\begin{equation}\label{101}
(\rho+p)(C^2C')^{\cdot}-\frac{\Omega}{32\pi}\frac{C}{C'}\overset{S}{=}0.
\end{equation}
The solution of this equation yields the surface $C_s$ which divides
the fluid into two regions with a positive and negative pressure
gradients respectively in the inner and outer sides of the surface.

\section{Summary and Conclusions}

The most important feature of $f(R)$ gravity is concerned with the
issue of DE. In this paper, we have studied the effects of $f(R)$
DE on the dynamics of a self-gravitating spherically symmetric
star. The star is made up of viscous anisotropic fluid
distribution and dissipating energy in the form of shearing
viscosity. This work extends and generalizes our recent paper
\cite{we2}. As a special case, the Skripkin model is studied which
is based on the simplest conditions of isotropic pressure and
constant energy density.

This study is devoted to the expansionfree evolution of the fluid
collapse resulting in the formation of a vacuum cavity within the
fluid distribution. We have developed junction conditions on two
hypersurfaces. One is the $\Sigma^{(e)}$, separating the fluid
distribution from the Schwarzschild spacetime while the other is
$\Sigma^{(i)}$, the boundary of internal cavity within which we
have Minkowski spacetime.

It is found that DE arising from the curvature fluid affects the
whole dynamics of the gravitational collapse due to its repulsive
effect. For example, it decreases the rate of change of mass of
collapsing sphere with respect to time and adjacent surfaces.  The
comparison with the corresponding GR yields more general results and
hence provides extra degree of freedom. For example, in Skripkin
model, the comparison with \cite{H1},\cite{H2} is given as follows
\begin{itemize}
\item In GR, implication of Skripkin conditions on the
dynamical equations yields expansionfree evolution, i.e.,
$\Theta=0$. However, no such relation is obtained in $f(R)$ gravity.

\item In GR, isotropic pressure with nondissipation
leaves the energy density as only a function of radial coordinate.
Here, Eq.(\ref{53}) does not imply that energy density is time
independent due to presence of curvature fluid (even in the
simplest case of constant scalar curvature).

\item Matching of results on the boundary surface $\Sigma^{(e)}$ gives
a non-zero constant value (based on constant scalar curvature) to
Skripkin energy density $\rho_0$ while it becomes zero in GR.

\end{itemize}
It it worth mentioning here that Skripkin model is completely
consistent with the junction conditions. Moreover, the
expansionfree models with the appearance of a vacuum cavity may be
used to study the formation of voids at cosmological scales
\cite{liddle} for any type of fluid.

\vspace{1.0cm}

{\bf Acknowledgment}

\vspace{0.25cm}

We would like to thank the Higher Education Commission, Islamabad,
Pakistan for its financial support through the {\it Indigenous Ph.D.
5000 Fellowship Program Batch-III}.

\end{document}